\documentclass[preprintnumbers, amsmath, amssymb, floatfix, twocolumn]{revtex4}
\usepackage{graphicx}
\usepackage{dcolumn}
\usepackage{bm}
\usepackage{color}%
\usepackage[update,prepend]{epstopdf}

\begin{document}

\title{Quantum oscillations in heavy fermion compound YbPtBi}
\author{E. Mun$^{1, 2}$, S. L. Bud'ko$^{1}$, Y. Lee$^{1}$, C. Martin$^{1, *}$, M. A. Tanatar$^{1}$, R. Prozorov$^{1}$, P. C. Canfield$^{1}$}
\affiliation{$^{1}$Ames Laboratory US DOE and Department of Physics and Astronomy, Iowa State University, Ames, IA 50011, USA}
\affiliation{$^{2}$Department of Physics, Simon Fraser University, Burnaby, BC V5A 1S6, Canada}


\begin{abstract}
We present quantum oscillations observed in the heavy fermion compound YbPtBi in magnetic fields far beyond its field-tuned, quantum critical point. Quantum oscillations are observed in magnetic fields as low as 60\,kOe at 60 mK and up to temperatures as high as 3\,K, which confirms the very high quality of the samples as well as the small effective mass of conduction carriers far from the  quantum critical point. Although the electronic specific heat coefficient of YbPtBi reaches $\sim$ 7.4 J/mol K$^{2}$ in zero field, which is one of the highest effective mass value among heavy fermion systems, it is suppressed quickly by applied magnetic field. The quantum oscillations were used to extract the quasiparticle effective masses of the order of the bare electron mass, which is consistent with the behavior observed in specific heat measurements. Such a small effective masses at high fields can be understood by considering the suppression of Kondo screening. 
\end{abstract}


\maketitle

\section{Introduction}

Heavy fermion (HF) compounds have provided some of the clearest evidence for quantum phase transitions \cite{Lohneysen2007}. A large effective mass introduces a characteristic low energy scale that can easily be tuned by non-thermal control parameters, making HF systems desirable for studying quantum criticality. An antiferromagnetic (AFM) quantum critical point (QCP) has been explored in many HF systems \cite{Stewart2001, Stewart2006}. However, a magnetic field-tuned QCP in Yb-based materials has been limited to only few cases, among stoichiometric compounds, in particular tetragonal YbRh$_{2}$Si$_{2}$ \cite{Gegenwart2002} and hexagonal YbAgGe \cite{Budko2004}. Recently, the HF compound YbPtBi attracted attention because of a quantum phase transition driven by magnetic field \cite{Mun2013}, in particular providing relatively easy access to the field-tuned QCP as well as the appearance of a new phase between the ordered (AFM) state and paramagnetic Fermi liquid state.

YbPtBi has face-centered cubic (F$\bar{4}$3m) structure and manifests AFM ordering below $T_{N}$ = 0.4 K with an electronic specific heat coefficient of $\gamma$ $\sim$ 7.4 J/mol K$^{2}$ at zero field \cite{Mun2013, Fisk1991}. Recent neutron scattering experiments characterized the averaged ordered moment of $\sim$ 0.8 $\mu_{B}$/Yb with the AFM propagation vector $\tau$ = (1/2, 1/2, 1/2) \cite{Ueland2014}. A field-tuned QCP is readily accessible by suppressing AFM ordering below $T_{N}$ = 0.4 K by application of external magnetic field of $H_{c}$ $\sim$ 4 kOe \cite{Mun2013}. The temperature dependence of the resistivity indicates a recovery of the Fermi liquid (FL) state for $H^{*}$ $\gtrsim$ 7.8 kOe. In addition, of particular interest in YbPtBi is the fact that a new phase appears in the proximity of the QCP. In this new phase, strange metallic metallic behavior (non-Fermi liquid) manifests and separates the AFM state and FL regime. A crossover scale has been found in thermodynamic and transport measurements \cite{Mun2013}, where the crossover line shows a tendency of converging toward to $H^{*}$ $\sim$ 7.8 kOe in the zero temperature limit. A similar phase near QCP and the crossover line have been observed in YbAgGe \cite{Budko2005, Mun2010} and Ge-doped YbRh$_{2}$Si$_{2}$ \cite{Custers2010}.

Although many examples of QCP have been reported in HF materials \cite{Stewart2001, Stewart2006}, theoretical classifications have not been firmly established yet. Unlike the traditional picture (e. g. for a spin density wave) of approaching a quantum criticality \cite{Hertz1976, Millis1993, Moriya1995}, where the heavy quasiparticles survive near the QCP,  the unconventional way (local quantum criticality) to explain quantum phase transition suggests a destruction of Kondo screening of the $f$-electrons \cite{Si2001, Si2003, Coleman2007}, involving Fermi surface reconstruction, where the heavy quasiparticles decompose into the conduction sea and local magnetic moments. If the Fermi surfaces have different shapes or are completely reconstructed across the QCP, then a Lifshitz transition, associated with a reconstruction of the Fermi surface, must separate the two phases between ordered and disordered state. In general, because most of the HF compounds have multiple Fermi surfaces, it is difficult to probe such a Lifshitz transition at extremely low temperatures. Transport measurements can address this issue clearly for a system with a single band, however this may not be true for a multiband systems. In metallic systems, the interpretation of physical quantities requires careful consideration of the Fermi surface, especially at the QCP.

In this paper, we report the determination of the high-field stabilized Fermi surfaces for YbPtBi inferred from quantum oscillations and band structure calculations. Until now, there was no direct experimental information about the size or shape of the Fermi surface in YbPtBi. By producing high quality single crystals, we are able to infer Fermi surfaces by detecting quantum oscillations down to 60 mK, but for only above $\sim$ 60 kOe, in resistivity measurements.

In principle, the Fermi surface can be studied by comparing the quantum oscillation measurements with electronic structure calculations. However, quantum oscillation data in most HF compounds is likely to be incomplete mainly due to the experimental difficulties. This is a standing problem in HF physics: in order to detect the heavier effective masses, higher magnetic fields and extremely lower temperatures (order of $\sim$ mK) are needed, however the mass enhancement can be suppressed due to the application of these larger magnetic fields. Further, when there is a 4$f$ electron contribution to the Fermi volume, band structure calculations can't accurately predict the Fermi surface topology in detail and an estimate of the density of states. Therefore, in this study, we focused data analysis on the high field-paramagnetic regime and compared it to the band structure calculations in zero field for simple, trivalent Yb. We confirm the multiband nature of this system, but are unable to identify the heavy Fermi surface. In addition, we compare our results to the family of $R$PtBi, where quantum oscillations have been observed for $R$ = Y, La, Ce, and Nd \cite{Butch2011, Goll2002, Wosnitza2006, Morelli1996}.

\section{Experimantal}

Single crystals of YbPtBi were grown out of a ternary melt with excess of Bi as has been described in earlier reports \cite{Canfield1991, Canfield1992}. Four-probe ac resistivity ($f$ = 16 Hz) measurements were performed in a Oxford dilution refrigerator, where the samples were attached to the cold state using GE-varnish. Pt wires were used as leads and attached to the sample with Epotek H20E silver epoxy. The magnetic field (\textbf{H}) was applied along the [100] and [111] direction, and the electric current (\textbf{I}) across the sample was applied perpendicular to \textbf{H}: \textbf{H}  $\perp$ \textbf{I}. Details of experimental conditions are given in Ref. \cite{Mun2013}.

In order to compare the experimental observations of Shubnikov-de Haas frequencies to the topology of the Fermi surfaces, we calculated the zero field band structure of paramagnetic, trivalent, YbPtBi. For the Fermi surface calculation, we have used a full-potential Linear Augmented Plane Wave (fp-LAPW) \cite{Blaha2001} method with a local density functional \cite{Perdew1992}. The structure data was taken from reported experimental results \cite{Robinson1994}. To obtain the self consistent charge density we chose 1204 $k$-points in the irreducible Brillouin zone and set $R_{MT}\cdot K_{max}$ to 9.0, where $R_{MT}$ is the smallest muffin-tin radius and $K_{max}$ is the plane-wave cutoff. We used muffin-tin radii 2.5 for all Yb, Bi, and Pt atoms. The calculation was iterated with 0.0001 electrons of charge and 0.01 mRy of total energy convergence criteria. Although there was a discussion about 4$f$ electron pinning at the Fermi energy \cite{Oppeneer1997} and we were aware that Fermi surface is quite different under 4$f$ electrons influence \cite{McMullan1992} we treated 4$f$ electrons as core-electrons since we were interested in the high magnetic field, paramagnetic state. To obtain SdH frequencies we calculated 2-dimensional Fermi surfaces and integrated the Fermi surface area. We chose planes which were perpendicular to $k_{z}$-axis and had 0.01 (2$\pi/a$) interval. Each plane (-1\,$\leq$\,$k_{x}$\,,\,$k_{y}$\,$\leq$\,1) was divided with 100$\times$100 mesh. For a 3-dimensional Fermi surface, we used 2300 $k$-points in the irreducible Brillounin zone and a XcrysDen graphic program \cite{XcrysDen}.

\section{Result}

\begin{figure}
\centering
\includegraphics[width=1\linewidth]{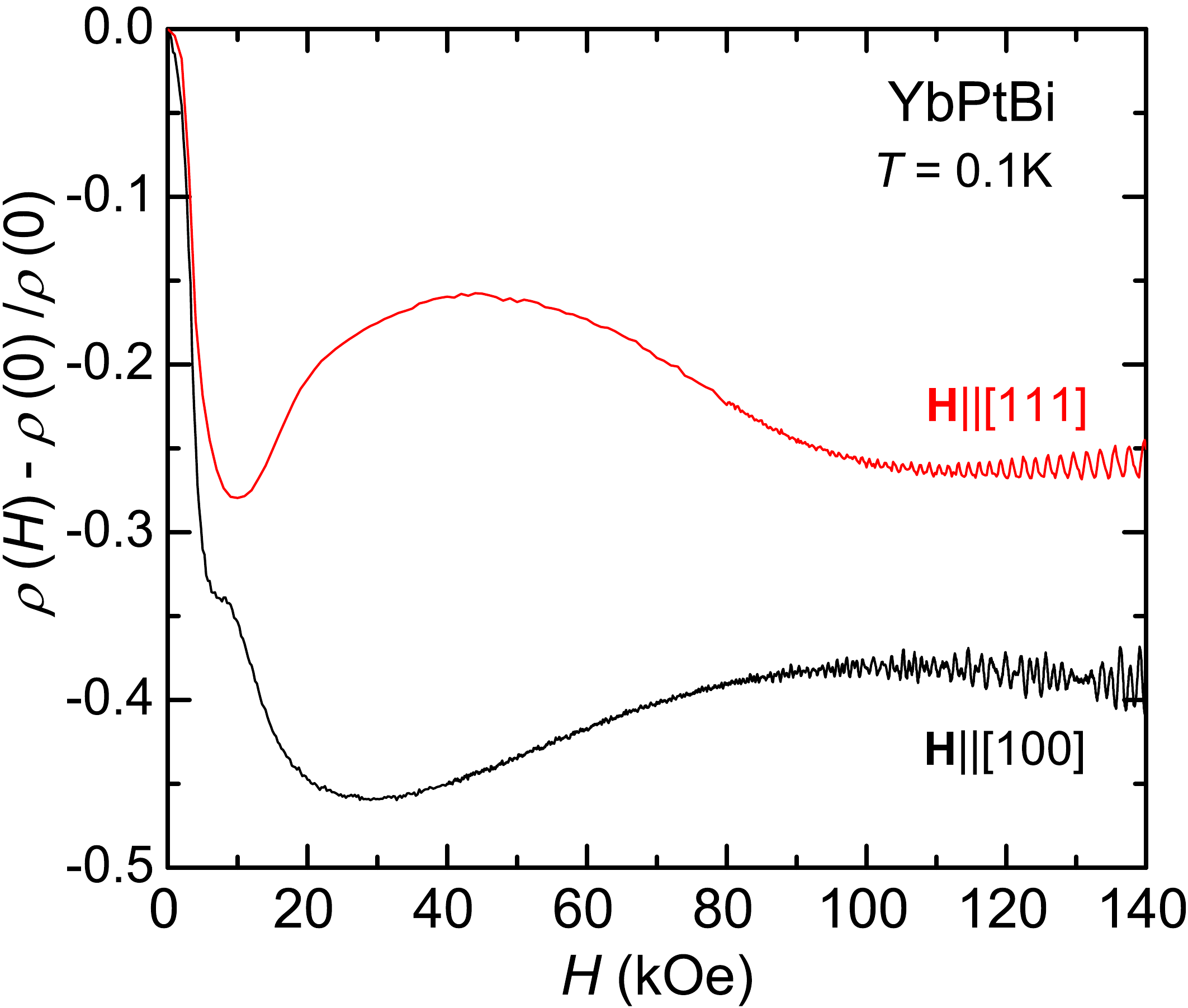}
\caption[Magnetoresistance of YbPtBi at high magnetic fields]{Magnetoresistance (MR), plotted as $[\rho(H)-\rho(0)]/\rho(0)$ vs. $H$, of YbPtBi at $T$ = 0.1\,K along \textbf{H}\,$\parallel$\,[100] and \textbf{H}\,$\parallel$\,[111]. At high magnetic fields quantum oscillations are discernible for both curves.}
\label{YbPtBiSdH1}%
\end{figure}%

Shubnikov-de Haas (SdH) quantum oscillations have been observed throughout the magnetoresistance (MR) measurements at low temperatures and high magnetic fields. Figure \ref{YbPtBiSdH1} shows the MR at $T$ = 0.1\,K for magnetic field applied along [100] and [111] directions. At high magnetic fields a broad local extrema in MR is observed for both magnetic field directions. This behavior may be due to the change of scattering processes with CEF levels, or it may be the oscillatory component corresponding to extremely small Fermi surface area in which the extremely small frequency has been observed for NdPtBi \cite{Morelli1996} in the paramagnetic state. For YbPtBi though this is not likely to be the case because the frequency is so small that it would have an amplitude that would make it hard to observe in SdH measurements. What is unambiguous are the clear quantum oscillations at high magnetic fields.

\begin{figure*}
\centering
\includegraphics[width=0.5\linewidth]{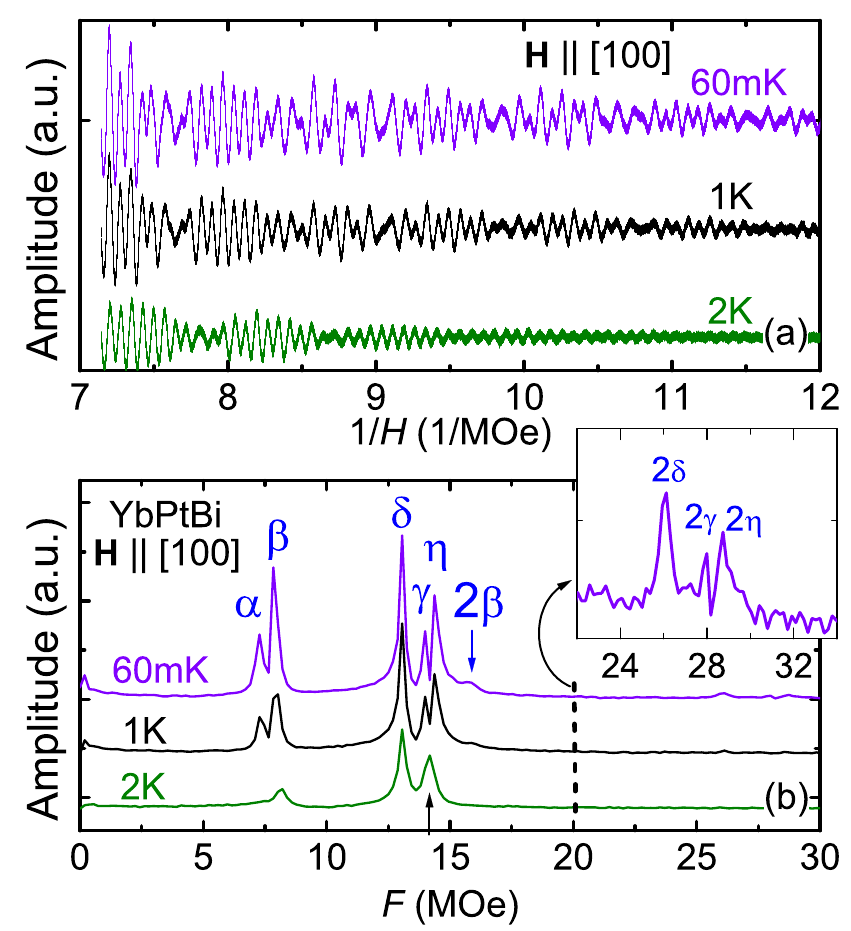}\includegraphics[width=0.5\linewidth]{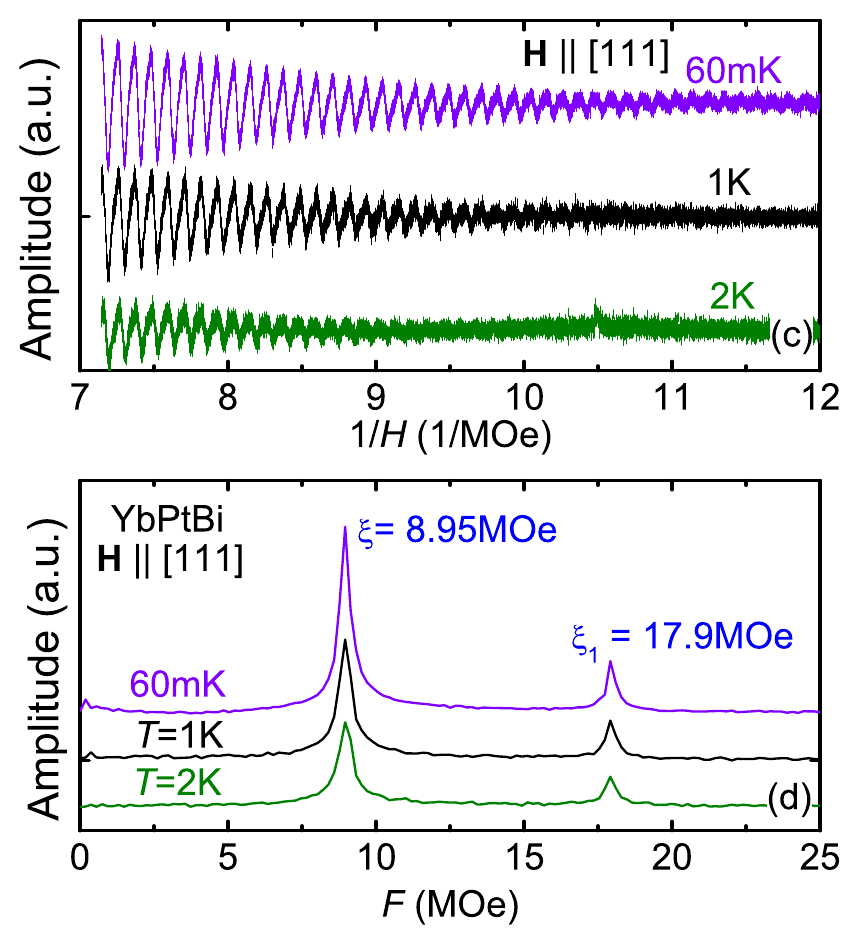}
\caption[Quantum oscillations observed in electrical resistivity measurements]{SdH of YbPtBi at $T$ = 0.06, 1, and 2\,K, plotted after subtracting the background MR, for (a) \textbf{H}\,$\parallel$\,[100] and (c) \textbf{H}\,$\parallel$\,[111]. FFT spectra of SdH data at $T$ = 0.06, 1, and 2\,K for (b) \textbf{H}\,$\parallel$\,[100] and (d) \textbf{H}\,$\parallel$\,[111].}
\label{YbPtBiSdH2}%
\end{figure*}%

In Figs. \ref{YbPtBiSdH2} (a) and (c) typical SdH data sets for YbPtBi, after subtracting the background contributions, are displayed as a function of 1/$H$ at selected temperatures. The amplitude of the oscillations decreases as temperature increases. Since the signals are comprised of a superposition of several oscillatory components, the data are most easily visualized by using a fast Fourier transform (FFT) as shown in Figs. \ref{YbPtBiSdH2} (b) for \textbf{H} $\parallel$ [100] and \ref{YbPtBiSdH2} (d) for \textbf{H} $\parallel$ [111]. The FFT spectra at $T$ = 0.06\,K show several frequencies, including, in the case of \textbf{H} $\parallel$ [100], second harmonics with very small amplitudes. The observed frequencies are summarized in Table \ref{YbPtBiSdHtable}.

\begin{table}
\caption[Frequencies and effective masses of SdH oscillation]{Frequencies $f$ and effective masses $m^{*}$ obtained from the SdH oscillations. $m_{e}$ is the bare electron mass.}
\label{YbPtBiSdHtable}%
\centering
\begin{tabular}{cccc}

 ~~~~~~~~  &   ~~~~~~~~    &  ~~~~  $f$ (MOe) ~~~~ &   ~~~~  $m^{*}/m_{e}$ ~~~~  \\ \hline
\textbf{H}\,$\parallel$\,[100] &      &  &   \\ 
   &$\alpha$      & ~~7.27   & 1.41  \\
   &$\beta$       & ~~7.83   & 1.59  \\
   &$\delta$      & 13.06    & 0.83 \\
   &$\gamma$      & 13.99    & 0.80 \\
   &$\eta$        & 14.37    & 0.97   \\
   &2$\beta$      & 15.67    &       \\
   &2$\delta$     & 26.13    &       \\
   &2$\gamma$     & 27.99    &       \\
   &2$\eta$       & 28.74    &       \\ \hline
\textbf{H}\,$\parallel$\,[111] &     &           &   \\
   &$\xi$         &  8.95    & 1.22    \\
   &$\xi_{1}$     & 17.90    & 0.49    \\  
\end{tabular}
\end{table}

Quantum oscillations are observed in magnetic fields as low as 60\,kOe at the lowest temperature measured and at temperatures as high as 3\,K, which further confirms the very high quality of the samples as well as the relatively small effective masses of the associated conduction electrons. The frequencies in FFT spectra do not shift with temperature and most of the first harmonics of the frequencies are clearly observed as high as 2\,K.

\begin{figure}%
\centering
\includegraphics[width=1\linewidth]{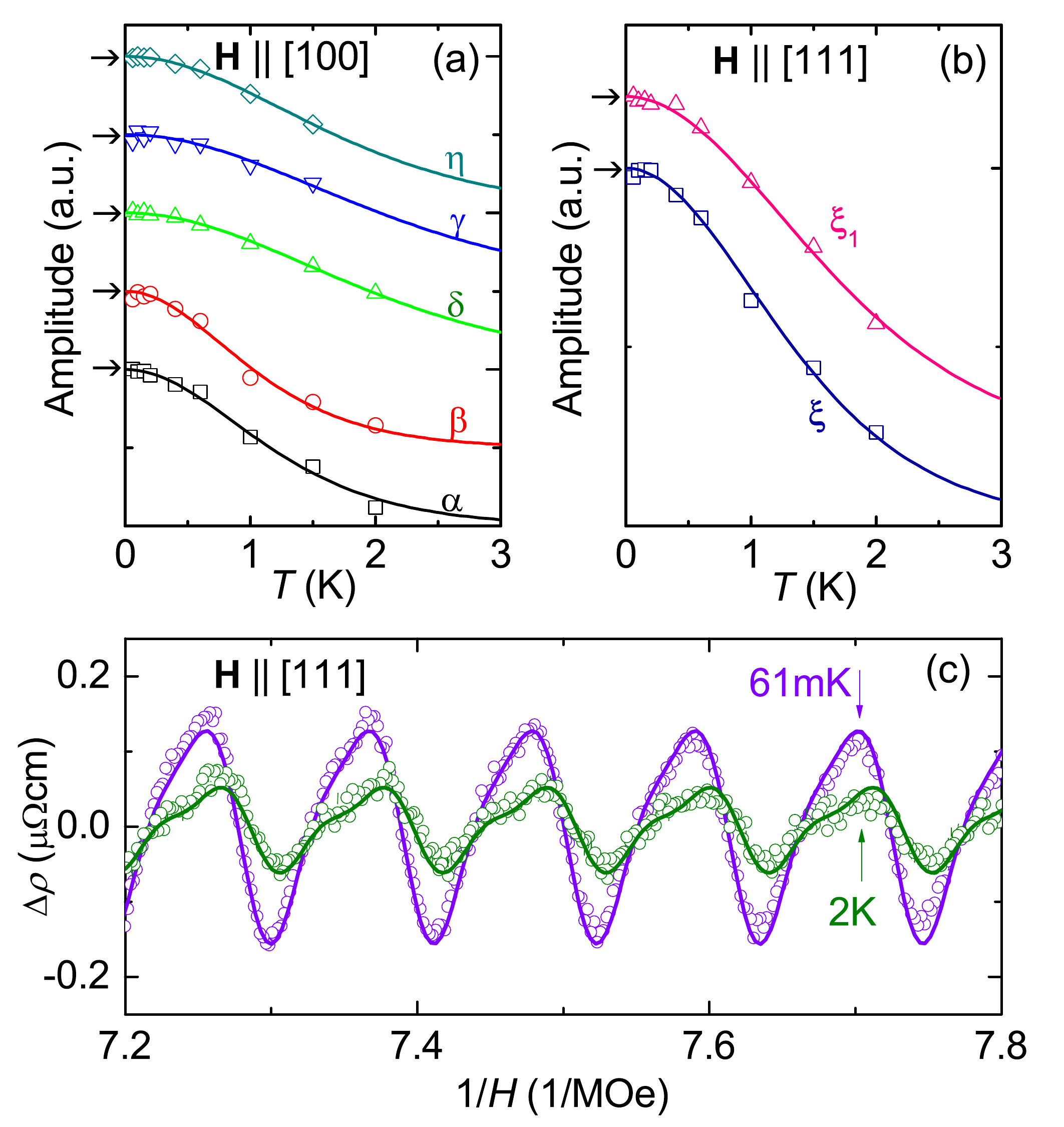}
\caption[Effective mass plot]{(a) and (b) Temperature dependence of the SdH amplitudes. Solid lines represent the fit curves to the Lifshitz-Kosevich (L-K) formula. All data and fit curves are normalized to 1, indicated by horizontal arrows, and shifted for clarity. (c) Resistivity along \textbf{H}\,$\parallel$\,[111] at $T$ = 0.06 and 1\,K, plotted as a function of 1/$H$ after subtracting the background MR, where the solid lines represent the fit curves based on the L-K formula with the frequency $\xi$ and $\xi_{1}$.}
\label{YbPtBiSdH4}%
\end{figure}

The oscillation amplitudes and the fit curves using the temperature reduction factor are plotted in Figs. \ref{YbPtBiSdH4} (a) and (b). The cyclotron effective masses, $m^{*}$, of the carriers from the various orbits were determined by fitting the temperature-dependent amplitude of the oscillations to the Lifshitz-Kosevich (L-K) formula \cite{Shoenberg1984} for each frequency. The calculated effective masses range from $m^{*}(\alpha)\,\sim$ 1.41\,$m_{e}$ to $m^{*}(\xi_{1})\,\sim$ 0.49\,$m_{e}$, where $m_{e}$ is the bare electron mass. The inferred effective masses are summarized in Table \ref{YbPtBiSdHtable}. We were not able to estimate the effective masses, associated with the second harmonic frequencies due to the small amplitude of the signals. Although the frequency $\xi_{1}$ is integer-multiple of $\xi$, $\xi_{1}$\,$\simeq$\,2$\xi$, it does not appear to be a higher harmonic of $\xi$ because of the inconsistent effective masses. In addition, if these frequencies are originating from the same extremal orbit, the phase difference between two frequencies can not be explained; the oscillation curves are generated by L-K formula with the phase term, $A_{1} \sin(2\pi \xi/H + \pi/1.95)$ + $A_{2} \sin(2\pi \xi_{1}/H - \pi/7.7)$, as shown in Fig. \ref{YbPtBiSdH4} (c). Therefore, $\xi_{1}$ really appears to be an independent frequency, coming from different extremal orbit. The frequency of the orbit $\eta$ is almost twice of the frequency of the $\alpha$, however these orbits are also apparently coming from different Fermi surfaces. If the orbit $\eta$ is the second harmonics of the $\alpha$, the oscillation amplitude of the $\eta$ should be smaller than that of $\alpha$, but the amplitudes of these frequencies are almost the same. Therefore, the orbit $\eta$ is not a second harmonic of $\alpha$. Note that the frequency, observed near 14\,MOe at 2\,K along \textbf{H}\,$\parallel$\,[100], seems to split from one component into two component of $\gamma$ and $\eta$ with decreasing temperature, as indicated by up arrow in Fig. \ref{YbPtBiSdH2} (b). At present it is not clear whether two frequencies of $\gamma$ and $\eta$ are originating from the same extremal orbit, thus it needs to be clarified by further detailed measurements.

\section{Discussion}

The low carrier density for YbPtBi, determined from Hall coefficient measurements \cite{Mun2013}, implies a Fermi surface occupying a small portion of the Brillouin zone, which is consistent with the results of our analysis of the quantum oscillations. The frequency of the quantum oscillations is proportional to the extremal cross-section, $A_{FS}$, of the Fermi surface; $f\,=\,(\hbar/2\pi e)A_{FS}$ \cite{Shoenberg1984}. In the high-field, paramagnetic region direct evidence for small Fermi surfaces comes from SdH measurements, where several small extremal orbits, implying a small portion of occupation of the Brillouin zone, are observed. Quantum oscillations have also been observed for LaPtBi and CePtBi \cite{Goll2002, Wosnitza2006} from the electrical resistivty measurements in pulsed magnetic fields up to 50\,Tesla. The oscillation frequencies for LaPtBi are approximately 10 times smaller, 0.65\,MOe for \textbf{H}\,$\parallel$\,[100] increasing to 0.95\,MOe for \textbf{H}\,$\parallel$\,[110], than for YbPtBi. In this family, SdH for YPtBi are observed up to 10 K with a frequency of 0.46 MOe \cite{Butch2011}. Note that a single SdH oscillation with $f$ = 0.74 T was inferred from magnetoresistiance measurements in NdPtBi \cite{Morelli1996}, where the observed frequency is two order of magnitude smaller than the frequencies observed for other compounds in this family. It is not clear whether the origin of the extraordinarily small frequency for NdPtBi is related to the broad local extrema observed in YbPtBi (Fig. \ref{YbPtBiSdH1}). For CePtBi the anomalous temperature dependence of SdH frequency, $f$ = 0.6\,MOe, was observed along \textbf{H}\,$\parallel$\,[100] and a very low SdH frequency of $\sim$\,0.2\,MOe, which is independent of temperature, was found along \textbf{H}\,$\parallel$\,[111]. In addition to the unusual temperature dependence of the SdH frequency for CePtBi, the disappearance of the oscillations was observed above about 25 T at which the magnetic field-induced band structure change was proposed \cite{Wosnitza2006}. 

Since the SdH frequencies for YbPtBi are not changed by temperature or magnetic field, within the temperature and magnetic field range of our measurements, such a band structure modification is not expected. The band calculations for LaPtBi \cite{Oguchi2001} and CePtBi \cite{Goll2002}, assuming localized 4$f$ states, were found to be semimetals. In these calculations, two hole-like Fermi surface sheets are found around zone center, which are similar to the measured angular dependence of the Fermi surface cross-section area of LaPtBi. A number of small electron-like pockets are also predicted in the band calculations. The effective masses for both LaPtBi and CePtBi have been estimated to be $\sim$ 0.3\,$m_{e}$ \cite{Wosnitza2006}, which is somewhat smaller than for YbPtBi. The observed trend of SdH frequencies suggested larger Fermi surface sheets for YbPtBi than LaPtBi, and these are consistent with earlier resistivity results of $R$PtBi \cite{Canfield1991}, where the resistivity varied from metallic (semimetallic) to small gap semiconductor when rare-earth changes from Lu to Nd \cite{Canfield1991}; $\rho(T)$ of LuPtBi decreases and $\rho(T)$ of NdPtBi increases as temperature decreases. The carrier density for LuPtBi is approximately two order of magnitude bigger than that for LaPtBi \cite{Mun2013}.

\begin{figure}%
\centering
\includegraphics[width=1\linewidth]{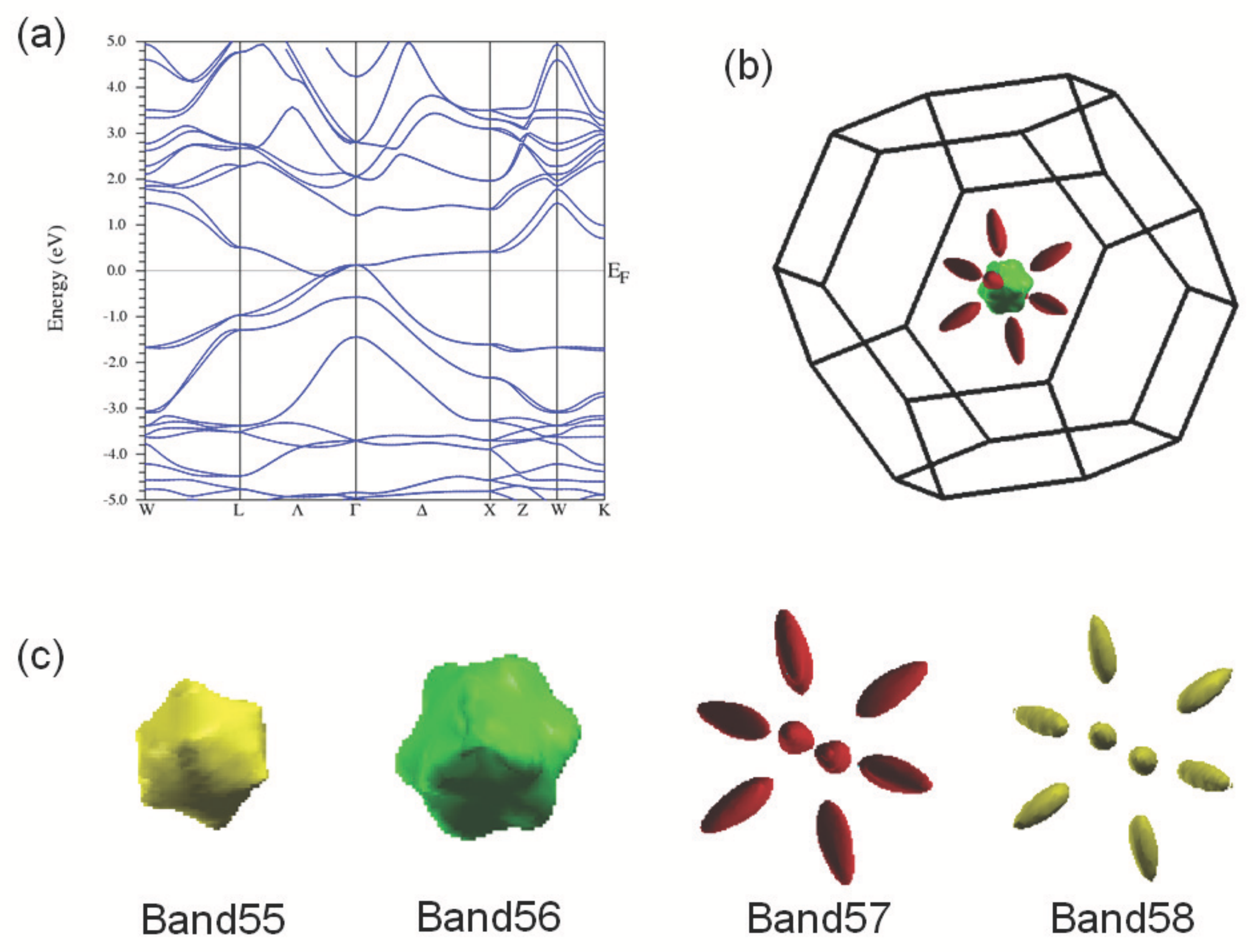}
\caption[Band structure calculations and Fermi surfaces for YbPtBi]{(a) Band structure of nonmagnetic YbPtBi, calculated for localized 4$f$
states. (b) Calculated Fermi surface in the fcc Brillouin zone. Two three dimensional pockets (bands 55 and 56), located in the zone center, are
surrounded by sixteen cigar shaped pockets (bands 57 and 58). (c) Enlarged Fermi surface for band 55, 56, 57, and 58.}
\label{YbPtBiBand}%
\end{figure}%

The results of band structure calculations are shown in Fig. \ref{YbPtBiBand} (a). The overall features are very similar to the results of LaPtBi, however with a larger Fermi surface occupation of the Brillouin zone. Around zone center two hole-like Fermi surface (band 55 and 56) are surrounded by sixteen small electron-like pockets (band 57 and 58). These calculated Fermi surfaces of YbPtBi are plotted in Fig. \ref{YbPtBiBand} (b) and (c). The SdH frequencies of these four bands are calculated to be 2.4, 3.5, 0.79, and 0.65\,MOe from the maximum area perpendicular to the $k_{z}$, where there are orbits very close to 2.4 and 3.5\,MOe due to the 3-dimensional shape of the Fermi surfaces at the zone center. These values are much larger than the predicted value for LaPtBi and CePtBi \cite{Goll2002, Wosnitza2006}, however four times smaller than the frequencies determined from our experimental results.

The Fermi surfaces of YbPtBi are highly sensitive to the 4$f$ electron contributions as predicted in Ref. \cite{McMullan1992}. When the 4$f$ electrons are included in the band calculations, the six hole-like pockets are located zone center in which the predicted frequencies range from $\sim$ 27 to $\sim$ 164\,MOe \cite{McMullan1992}, which is much higher than the experimental observations. So treating 4$f$ electrons as included in core levels appears to be reasonable. If the Fermi level is shifted to lower energy, the experimentally observed frequencies can be matched to the hole-like pockets at the zone center, whereas the electron-like pocket surrounding the zone center will not be detected. As a conjecture, the effective masses and frequencies of the orbit $\alpha$ and $\beta$ along [100] direction, linked to the $\xi$ along [111] direction, are almost the same, expected that these two orbits are came from the band 55. Similarly the orbits $\delta$, $\gamma$, and $\eta$, connected to the $\xi_{1}$, all came from the 3-dimensional shape of the band 56. Without an angular dependence of the SdH measurements, the Fermi surface topology can not be determined unambiguously and further theoretical work is needed to unravel the discrepancies in the precise extremal orbit sizes. 

Since we have observed only small effective masses for YbPtBi, it is expected that the hybridization between 4$f$ and conduction electrons has been suppressed for these high magnetic fields. This is consistent with the specific heat results; the enormous value of $\gamma$ $\sim$ 7.4 J/mol$\cdot$K$^{2}$ for $H$ = 0 is suppressed to $\gamma$ $\sim$ 0.15 J/mol$\cdot$K$^{2}$ for $H$ = 50\,kOe and would extrapolate to $\sim$ 0.030 J/mol$\cdot$K$^{2}$ at 140 kOe (using date from \cite{Mun2013}). Note that if there are still significant 4$f$ electron contributions in this magnetic field range, up to 140\,kOe, it would require much lower temperatures to observe heavy electrons in SdH measurements. This is a standing problem in HF physics, in order to detect the heavier effective masses, higher magnetic fields and very low temperatures are needed, however the mass enhancement can be suppressed due to the application of these larger magnetic fields. Thus, lower measurement temperatures and crystals with extremely low scattering in terms of Dingle temperature are necessary to detect heavier effective mass of carriers. Measuring de Haas-van Alphen (dHvA) oscillations as a complementary to SdH oscillations may be another experimental approach since oscillation amplitudes have different dependence of $m^{*}$ in dHvA and SdH. However, this task can be challenging due to high paramagnetic background signal.

\section{summary}
In summary, we have presented Shubnikov-de Haas quantum oscillations detected in YbPtBi, which is the first report on quantum oscillations since the discovery of its heavy fermion behavior in 1991. The band structure calculations for the Fermi surface are also presented, where treating 4$f$ electrons as included in core levels appears to be important in high field regime far beyond the quantum critical point. Comparison is made between the high field oscillations and zero field band structure calculations, allowed us to infer Fermi surfaces only in the paramagnetic state. The current study clearly shows that i) small oscillation frequencies suggest a low carrier density, consistent with Hall effect measurements, ii) multiple oscillation frequencies imply a multiband nature, and iii) small effective masses at high fields indicate a suppression of Kondo screening, consistent with specific heat results. Although it was not possible to infer heavy fermion state, the current study may, in future, give useful information about the nature of the quantum criticality when the Fermi volume at low fields are available.

\begin{acknowledgments}
This work was supported by the U.S. Department of Energy, Office of Basic Energy Science, Division of Materials Sciences and Engineering. The research was performed at the Ames Laboratory. Ames Laboratory is operated for the U.S. Department of Energy by Iowa State University under Contract No. DE-AC02-07CH11358. The work at Simon Fraser University was supported by Natural Sciences and Engineering Research Council of Canada. 
\end{acknowledgments}

\clearpage

$^{*}$Current address; Ramapo College of New Jersey, Mahwah, New Jersey 07430, USA.

\end{document}